\documentclass[]{interact}
\usepackage{fix-cm}

\usepackage{epstopdf}
\usepackage{subfigure}

\usepackage{natbib}
\bibpunct[, ]{(}{)}{;}{a}{}{,}
\renewcommand\bibfont{\fontsize{10}{12}\selectfont}

\usepackage{tikz}
\usepackage{multirow} 
\usepackage{pgfplots}
\usepackage{algorithm}
\usepackage{algpseudocode}
\pgfplotsset{compat=1.17} 

\theoremstyle{plain}

\theoremstyle{definition}

\theoremstyle{remark}

\begin{document}

\title{An Asynchronous Many-Task Algorithm for Unstructured $S_{N}$ Transport on Shared Memory Systems}

\author{
\name{Alex Elwood\textsuperscript{a}\thanks{CONTACT Alex Elwood. Email: alex.elwood@bristol.ac.uk \\ UK Ministry of Defence © Crown Owned Copyright 2025/AWE.}, Tom Deakin\textsuperscript{a}, Justin Lovegrove\textsuperscript{b} and Chris Nelson\textsuperscript{c}}
\affil{\textsuperscript{a}High Performance Computing Research Group, School of Computer Science, University of Bristol, Bristol, UK; \textsuperscript{b}Computational Physics Group, AWE, Aldermaston, UK;  \textsuperscript{c}High Performance Computing Group, AWE, Aldermaston, UK}
}

\maketitle

\begin{abstract}
Discrete ordinates $S_N$ transport solvers on unstructured meshes pose a challenge to scale due to complex data dependencies, memory access patterns and a high-dimensional domain. In this paper, we review the performance bottlenecks within the shared memory parallelization scheme of an existing transport solver on modern many-core architectures with high core counts. With this analysis, we then survey the performance of this solver across a variety of compute hardware. We then present a new Asynchronous Many-Task (AMT) algorithm for shared memory parallelism, present results showing an increase in computational performance over the existing method, and evaluate why performance is improved.
\end{abstract}

\begin{keywords}
Neutral particle transport; Asynchronous Many-Task; Unstructured
\end{keywords}

\section{Introduction}

Neutral particle transport methods aim to simulate the interaction between neutral particles (such as neutrons, photos, etc.) and a material. Simulations must discretize the spatial, angular, energy and time domains in order to numerically compute a solution for angular flux. Due to the speed, size and interaction infrequency of the simulated particles, an extremely fine-grained discretization is often required across some or all dimensions of the domain. As such, solving this simulation requires huge amounts of computational resources to achieve sufficient accuracy and resolution of the angular flux. It is essential that these simulations efficiently leverage the compute hardware to decrease runtime and increase the resolution that can be tractably achieved.

Strategies to parallelize discrete ordinates transport sweeps on 2D and 3D structured meshes were proposed in earlier works for a message passing paradigm. For a 3D mesh, the spatial domain is decomposed across two of the dimensions where an individual processor core will compute the solve for a \textit{pencil} and communicate data as necessary to other cores, this is known as a KBA sweep \citep{Koch1992, Baker1998}. Later work then adapted the KBA sweep for unstructured grids \citep{Nemanic1999, Pautz2001, Plimpton2000}. Primarily, these works focus on discussing optimizations to the algorithms regarding prioritization of tasks and efficiently adapting the KBA spatial domain decomposition for unstructured grids. This work is then updated for more recent hardware and achieves a greater parallel efficiency than previous works \citep{Baker2017}. 

More recent works on semi-structured meshes have managed to achieve even greater scaling than methods using fully-unstructured meshes \citep{Adams2020, Vermaak2020}. These techniques can begin by taking unstructured grids and converting them to semi-structured, however, this significantly increases the number of mesh elements and therefore memory requirements. These will not be considered in this paper as we are concerned purely with unstructured meshes, although our proposed techniques can also be applied to the semi-structured case.

Trends in modern processor architectures have recently favoured increasing the number of cores over increasing the speed of each core. Specifically, the highest end Intel server CPUs have seen a more than doubling of the number of cores over the past decade and AMD server CPUs have seen an eight times increase in core count. Therefore, it is likely that the bottlenecks constraining the performance of some programs will have changed and the importance of ensuring our programs scale well is all the more pertinent. As such we must revisit the algorithms which our existing applications rely in order to understand how they perform on modern processors with high core counts.

In this paper we evaluate the performance limiting characteristics of an existing neutral particle transport solver parallelization scheme on modern processors featuring a higher core count. Secondly, we describe a new algorithm for parallelizing transport solvers using an Asynchronous Many-Task (AMT) paradigm. This evaluation and new method will be applied to the UnSNAP proxy application (mini-app) and implemented using OpenMP \citep{Deakin2018}.

Other works have also proposed asynchronous approaches to computing transport sweeps but have some fundamental differences. Specifically the works of \citet{Yan2013} and later \citet{Yan2023} propose solvers which use a producer thread to handle distributed communication and populate shared work queues for workers to consume tasks from. Since these papers, CPU core counts have continued to rise, causing the relationship of 1 producer to N consumers to be fundamentally unscalable for large enough N. 

Section \ref{sec:review} describes the current parallelization scheme used by UnSNAP, detailing how it is parallelized within a single compute node and across multiple compute nodes. The section then presents contemporary benchmark results for this existing scheme where the performance constraining limitations are discussed to establish a new baseline performance on the latest computer processors. Section \ref{sec:amt} presents our new parallel algorithm, including the design and implementation decisions, a comparison with the previous method and an analysis of the performance limiting factors. The paper concludes in section \ref{sec:conclusion} where outcomes and future work are discussed.

\section{Reviewing Bulk Synchronous Parallelization of UnSNAP}
\label{sec:review}

UnSNAP is a mini-app which uses an $S_N$ discrete ordinates method and a discontinuous Galerkin Finite Element Method (FEM) in the spatial domain to solve the time independent neutral particle transport equation on unstructured meshes of hexahedral elements \citep{Deakin2018} . It is an adaptation of the SNAP mini-app, a neutral particle transport solver for structured meshes using a Finite Difference method \citep{Zerr2013}. 

\begin{subequations} \label{eq:transport}
\begin{align}
\hat\Omega\cdot\overset\rightarrow\nabla\psi(\overset\rightarrow r, \hat\Omega, E)+ \label{eq:transport1}\\
\sigma(\overset\rightarrow r, E)\psi(\overset\rightarrow r, \hat\Omega, E)= \label{eq:transport2}\\
q_{ex}(\overset\rightarrow r, \hat\Omega, E)+ \label{eq:transport3}\\
\int dE' \int d\Omega'\sigma_s (\overset\rightarrow r, E'\rightarrow E,\hat\Omega'\cdot\hat\Omega)\psi(\overset\rightarrow r,\hat\Omega',E') \label{eq:transport4}
\end{align}
\end{subequations}

The goal of the mini-app is to numerically solve for the angular flux $\psi(\overset\rightarrow r, \hat\Omega, E)$ across the domain of the function, as defined by equation \ref{eq:transport}. To achieve this, the mini-app discretizes over the spatial domain, $\overset\rightarrow r$, using an unstructured mesh of hexahedral finite elements, divides the energy domain into a discrete set of energy groups, $E$, and discretizes the angular domain, $\hat\Omega$, using an $S_N$ quadrature set. These angles are grouped into octants where the direction vector of each angle will have the same signs as all other angles within the octant. The external source $q_{ex}(\overset\rightarrow r, \hat\Omega, E)$ is the fixed change in flux originating outside the physics modelled by the equation. Cross sections $\sigma(\overset\rightarrow r, E)$ describe how particles will interact with the material under certain types of reaction. In this case, the external source and cross sections are both taken from arrays of constant synthetic data.
 
The mini-app uses synthetic data which is an unstructured orthogonal grid of hexahedrons which is then twisted along one axis. As such, it is not possible to study any aspects of the transport solver which affect the convergence rate as that is dependent on the dataset. Specifically, there cannot be discussion of the number of ranks within the problem or evaluating different spatial decomposition schemes as UnSNAP uses pre-specified mesh sizes and domain decompositions. However, there is extreme value in discussing the performance of the parallel algorithms necessary to calculate an update to the angular flux as the numerical properties (such as convergence, order, etc.) of the solver are unchanged.

\subsection{Shared Memory Parallelism}
\label{sec:par-strat}

Solving for each value of angular flux requires a \textit{sweep} through the unstructured grid for every angle in each octant. As the sweep progresses through the grid, it will hit an element. This element is then solved using FEM for every energy group and the sweep progresses onto the next \textit{downwind} neighbors. These neighbors then have a data dependence on the previous \textit{upwind} element, as is the case for all subsequent downwind elements. This creates a complex structure of dependencies which can be represented as a Directed Acyclic Graph (DAG). Crucially, as mesh data is presented to the application at runtime, the application must be generalized to solve for any such mesh.

UnSNAP uses the method described by \citet{Deakin2019} to compute this sweep in parallel. A schedule is first precomputed before executing sweeps. To do this, the mesh is traversed for each angle and elements are placed into \textit{buckets}, which store elements within a level of the DAG, the \textit{t-level} (see Figure \ref{fig:schemes}). This simplifies the dependencies between the buckets of adjacent t-levels instead of between individual elements; elements within a bucket can be computed in parallel for all energy groups. UnSNAP implements this in OpenMP with a parallel loop across all elements and energy groups within a bucket. This is then repeated for all buckets, angles, and octants. UnSNAP vectorizes the loops within each FEM solve. This parallelism paradigm is known as Bulk Synchronous Parallel (BSP), characterized by separating work across processors, independently computing each work item, then synchronizing to communicate results. In UnSNAP this means that there is a synchronization barrier after every bucket (equivalent to t-level/wavefront). 

Additionally, UnSNAP has the option to compute angles within an octant simultaneously.\footnote{One intentional restriction within UnSNAP is not to compute octants in parallel due to possible optimizations when using reflective boundary conditions although the authors are aware of the optimizations this could potentially bring \citep{Adams2020}.} It does this by merging together buckets at the same t-level from all angles within an octant after they have been precomputed. The advantage is that it exposes more parallelism and reduces synchronization points as there are fewer buckets to loop over. However, this means that the values of scalar flux are accumulated in parallel as defined the $S_N$ discrete ordinates approximation given by equation \ref{eq:scalar-dis}, meaning data races may occur if multiple angles are computed simultaneously which must be prevented via atomic operations. The impact is that reading and writing of data is significantly slower when multiple threads try to access the same piece of data through atomic operations. 

\begin{subequations}  \label{eq:scalar}
	\begin{align}
		\phi(\overset\rightarrow r,E) = \int\psi(\overset\rightarrow r,\hat\Omega,E) d\hat\Omega \label{eq:scalar-con} \\
		\phi(\overset\rightarrow r,E) \approx \sum_n w_n\psi_n(\overset\rightarrow r,\hat\Omega,E) \label{eq:scalar-dis}
	\end{align}
\end{subequations}

The other major disadvantage is that computing multiple angles simultaneously requires the data for all those angles not just an individual angle. Table \ref{tab:arrays} summarizes all of the data arrays referenced within each FEM assemble and solve for an individual element. It shows that there are 6 arrays which have a domain over angle, each sweep in a parallel angle scheme would reference the entire array instead of just the array section for a given angle. This would increase the data required by lower level caches for these arrays, leaving less space for other data which contributes to poorer cache utilization. Previous literature has shown that performance models for sequential in angle schemes show greater scaling, however, simultaneous in angle schemes generally perform faster in practice \citep{Pautz2001}. 

\begin{table}[htbp]
	\tbl{Mesh and environmental data arrays accessed within each FEM Solve and the dimensions that they are defined across.}
	{\begin{tabular}{lcccccccc} \toprule
			\multirow{2}{*}{\textbf{Array}} & \multirow{2}{*}{\textbf{Symbol}} & \multicolumn{7}{c}{\textbf{Dimension}} \\
			& & Face & Moments & Material & Groups & Elements & Angles & Octant\\ \midrule
			\texttt{mu} & $\mu$ & & & & & & $\checkmark$ & $\checkmark$  \\
			\texttt{eta} & $\eta$ & & & & & & $\checkmark$ & $\checkmark$  \\
			\texttt{xi} & $\xi$ & & & & & & $\checkmark$ & $\checkmark$  \\
			\texttt{sigt} & $\sigma$ & & & $\checkmark$ & $\checkmark$ & & & \\
			\rule{0pt}{3ex}\texttt{fi\_dfj\_dx}\textsuperscript{a} & $f_i \cdot \frac{\partial f_j}{\partial x} \cdot \text{det}J$ & & & & & $\checkmark$ & & \\
			\rule{0pt}{3ex}\texttt{fi\_dfj\_dy} & $f_i \cdot \frac{\partial f_j}{\partial y} \cdot \text{det}J$ & & & & & $\checkmark$ & & \\
			\rule{0pt}{3ex}\texttt{fi\_dfj\_dz} & $f_i \cdot \frac{\partial f_j}{\partial z} \cdot \text{det}J$ & & & & & $\checkmark$ & & \\
			\rule{0pt}{3ex}\texttt{fi\_dfj} & $f_i \cdot f_j \cdot \text{det}J$ & & & & & $\checkmark$ & & \\
			\texttt{source} & $q_{ex}$ & & $\checkmark$ & & $\checkmark$ & $\checkmark$ & & \\
			\texttt{ec}\textsuperscript{b} & & & $\checkmark$ & & & & $\checkmark$ & $\checkmark$  \\
			\texttt{normals} & $n$ & $\checkmark$ & & & & $\checkmark$ & & \\
			\texttt{face\_nodes} & & $\checkmark$  & & & & & & \\
			\texttt{elem\_neighbors} & & $\checkmark$ & & & & $\checkmark$ & & \\
			\texttt{angular\_flux} & $\psi$ & & & & $\checkmark$ & $\checkmark$ & $\checkmark$ & $\checkmark$ \\
			\texttt{scalar\_flux} & $\phi$ & & & & $\checkmark$ & $\checkmark$ & & \\
			\texttt{scalar\_flux\_moments} & $\phi_{l}$ & & $\checkmark$ & & $\checkmark$ & $\checkmark$ & & \\
			\texttt{w} & $w_n$ & & & & & & $\checkmark$ & \\ \midrule
			\textbf{Count} & & 3 & 3 & 1 & 5 & 10 & 6 & 5 \\ \bottomrule
	\end{tabular}}
	\tabnote{\textsuperscript{a}Arrays beginning \texttt{fi} are the FEM basis functions.}
	\tabnote{\textsuperscript{b}\texttt{ec} is used to compute the flux moments.}
	\label{tab:arrays}
\end{table}

\subsection{Distributed Memory Parallelism}
\label{sec:distributed}

As typical problem sizes require significantly more memory and compute performance than a single node can accommodate, distributed memory parallelism must also be implemented to leverage multiple nodes in parallel. Therefore, systems require a scheme to decompose the mesh data across multiple compute nodes so each can solve an individual chunk of work before communicating results. UnSNAP has two options for a spatial decomposition where the mesh is divided either in 3-dimensions into smaller cubes or 2-dimensions into \textit{pencils} as described by \citep{Koch1992}.

This poses another implementation decision in how dependencies are communicated across node boundaries within the decomposed mesh. UnSNAP currently has two options for this. A full parallel sweep (FPS) always respects node boundary dependencies, leading to the least number of iterations required for convergence. However, this causes poor node utilization as they sit idle until sweep node boundary dependencies are communicated (known in this field as a start up cost). A parallel block jacobi (PBJ) method has each local mesh perform a sweep independently with dependencies only communicated after each sweep. This ensures high node utilization resulting in fast sweep iterations but increases the number of iterations required for convergence. Additionally, a method has been developed which uses \textit{chaotic iterations} (CPS), where boundary dependencies are asynchronously communicated as they are computed; UnSNAP does not implement CPS \citep{Garrett2018}. CPS potentially offers a middle ground between FPS and PBJ in terms of sweep iteration time and number of iterations to convergence.

\subsection{Numerical Results}

Following on from \citet{Deakin2020}, we recreate the benchmark test carried out to evaluate the performance of UnSNAP on a variety of microarchitectures but aim to update it with the latest processors. As is the trend in current processor technology, the new microarchitectures which we use exhibit many more cores, larger caches and higher memory bandwidth. A summary of the hardware used is displayed in table \ref{tab:compute-nodes}. Each benchmark was performed on a $16\times16\times16$ grid of spatial finite elements, totaling 4096 elements, with 10 energy groups and 32 angles per octant. We also choose the simultaneous angle scheme, and allow it to perform five inner and five outer iterations to ensure a long enough runtime for stable results. This is then run for first, second, and third FEM orders across all the listed hardware. The AMD Genoa and Intel Sapphire Rapids support Simultaneous Multithreading (SMT) we find that in most cases, enabling this gives a smaller performance improvement, as such this will be used when available. Where SMT is used we bind OpenMP threads to physical threads and otherwise we bind it to cores. All available cores are utilized within the node. As we perform this benchmark on a single shared memory compute node, the distributed memory parallelism methods described in section \ref{sec:distributed} are unused and will not be discussed further as we solve the entire problem on a single node.

\begin{table}[htbp]
\tbl{List of compute nodes utilised during benchmarking.}{    
    \begin{tabular}{ccccccc}
    \toprule
    \multirow{3}{*}{\textbf{Type}} & \multirow{3}{*}{\textbf{Architecture}} & \multirow{3}{*}{\textbf{Processor}} & \textbf{Cores/} &  \multicolumn{3}{c}{\textbf{Total}} \\
    & & & \textbf{SMs}\textsuperscript{a} &  \multicolumn{3}{c}{\textbf{Cache (MB)}} \\
    & & & & L1 & L2 & L3 \\
    \midrule
    \multirow{7}{*}{CPU} & Broadwell & Intel Xeon E5-2680 v4 & $2\times14$ & 0.9 & 3.6  & 35 \\ 
    & Sapphire Rapids  & Intel Xeon Gold 6430 & $2\times32$ & 5 & 128 & 120 \\
    & Milan & AMD EPYC 7713 & $2\times64$ & 8 & 64 & 512 \\
    & Genoa & AMD EPYC 9354 & $2\times32$ & 4 & 64 & 512 \\
    & Bergamo & AMD EPYC 9754 & 128 & 8 & 128 & 256 \\
    & Grace & NVIDIA GH200 & 72 & 9.2 & 72 & 114 \\
    & Grace & NVIDIA Grace Superchip & $2\times72$ & 18.4 & 144 & 288 \\
    \midrule
    \multirow{4}{*}{GPU} & Pascal & NVIDIA P100 & 56 & 1.3 & 4 & N/A \\ 
    & Ampere & NVIDIA A100 SXM4 & 120 & 6.8 & 40 & N/A \\
    & Hopper  & NVIDIA H100 PCIe & 114 & 14.3 & 50 & N/A \\
    & Hopper & NVIDIA GH200 & 132 & 16.5 & 50 & N/A \\
    \bottomrule
    \end{tabular}}
	\tabnote{\textsuperscript{a}Nodes with $2\times X$ cores are dual socket.}
    \label{tab:compute-nodes}
\end{table}

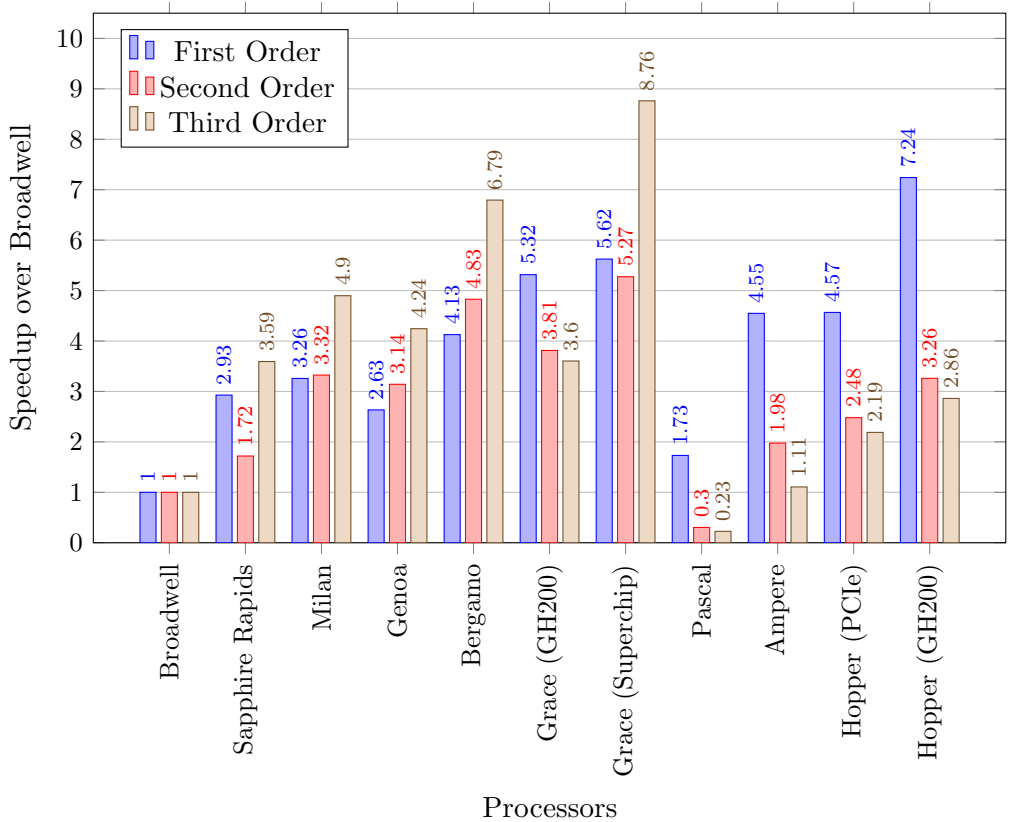
\begin{figure}
\centering
\begin{tikzpicture}
    \begin{axis}[
        ybar,
        bar width=6pt,
        symbolic x coords={Broadwell,
            Sapphire Rapids,
            Milan,
            Genoa,
            Bergamo,
            Grace (GH200),
            Grace (Superchip),
            Pascal,
            Ampere,
            Hopper (PCIe),
            Hopper (GH200)},
        xtick=data,
        ymajorgrids,
        ymin=0,ymax=10.5,
        ylabel={Speedup over Broadwell},
        xlabel={Processors},
        ytick={0, 1, 2, 3, 4, 5, 6, 7, 8, 9,10},
        legend pos=north west,
        nodes near coords,
        every node near coord/.append style={rotate=90, anchor=west, font={\footnotesize}},
        xticklabel style={rotate=90, anchor=east},
        x=1cm,
        width=\linewidth,
        height=0.6\linewidth,
        tick label style={font={\small}}
    ]
        \addplot+[
            ybar,
        ] table [x=Name, y=Speedup 1, col sep=comma] {benchmarks.csv};
        
        \addplot+[
            ybar,
        ] table [x=Name, y=Speedup 2, col sep=comma] {benchmarks.csv};
        
        \addplot+[
            ybar,
        ] table [x=Name, y=Speedup 3, col sep=comma] {benchmarks.csv};

        \legend{First Order, Second Order, Third Order}
    \end{axis}
\end{tikzpicture}
\caption{Benchmarks for a BSP implementation of UnSNAP across a variety of compute hardware.} 
 \label{fig:benchmarks}
\end{figure}

The results of this benchmark are shown in figure \ref{fig:benchmarks} and are displayed as speedups over Broadwell for the respective FEM order to match the original study. We observe that the speedups of the Pascal benchmark similarly match the results from the original study, giving confidence in close replication of the original study. We observe that GPUs generally perform better on lower order problems than higher order problems. However, the CUDA implementation for UnSNAP uses a different kernel for first order FEM solves which uses shared local memory for each assemble solve; this will be a contributing factor to the performance improvement. Nevertheless, GPU implementations ran on Ampere and Hopper (for both the PCIe and GH200 variants) show better performance on first order problems than most modern CPUs. This shows that there should still be interest in the performance of unstructured $S_N$ discrete ordinates solvers on GPUs.

\section{Asynchronous Many-Task Parallelization of UnSNAP}
\label{sec:amt}

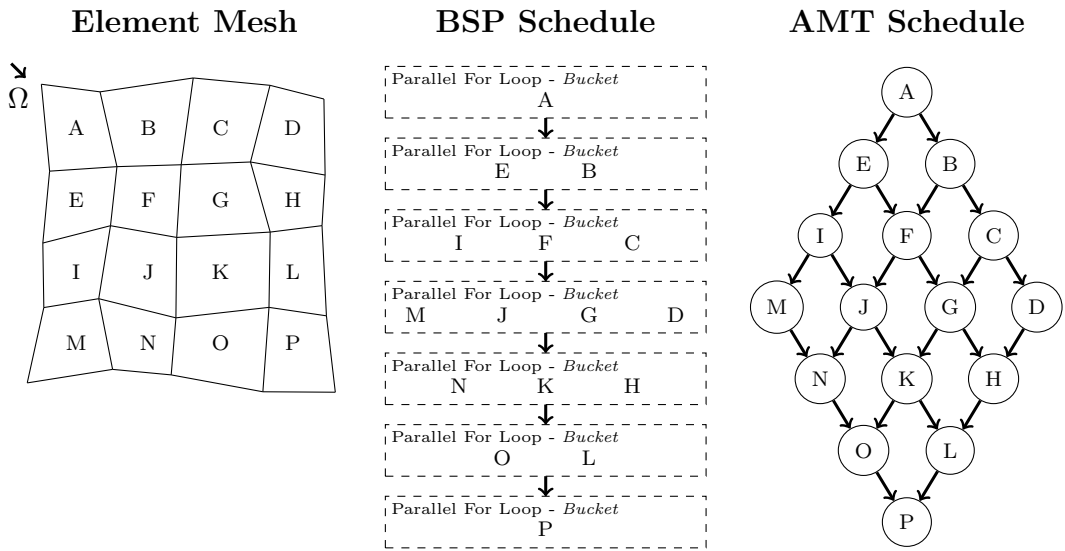
\begin{figure}[htbp]
    \begin{center}
    \begin{tikzpicture}[scale=0.95]
        \def\scale{1}
        \def\scaleT{0.6}
        \def\scaleBSP{0.6*\scaleT}
    
        \tikzset{
        	element/.style={font=\scriptsize},
            amtElement/.style={circle, minimum size = 1mm, draw=black, font=\scriptsize},
            bspElement/.style={font=\scriptsize},
        }

        \begin{scope}[xshift=-5.5cm, yshift=-0.5cm]
            \node[anchor=center] at (0.5, 1.5) {\large \textbf{Element Mesh}};            
            \draw[->, line width=0.4mm] (-1.9, 0.9) -- ++(0.2, -0.2);
            \node[anchor=north] at (-1.8, 0.7)  {$\Omega$};
            \node[element] (F-0-0) at (-1, 0) {A};
            \node[element] (F-1-0) at (0, 0) {B};
            \node[element] (F-2-0) at (1, 0) {C};
            \node[element] (F-3-0) at (2, 0) {D};
            \node[element] (F-0-1) at (-1, -1) {E};
            \node[element] (F-1-1) at (0, -1) {F};
            \node[element] (F-2-1) at (1, -1) {G};
            \node[element] (F-3-1) at (2, -1) {H};
            \node[element] (F-0-2) at (-1, -2) {I};
            \node[element] (F-1-2) at (0, -2) {J};
            \node[element] (F-2-2) at (1, -2) {K};
            \node[element] (F-3-2) at (2, -2) {L};
            \node[element] (F-0-3) at (-1, -3) {M};
            \node[element] (F-1-3) at (0, -3) {N};
            \node[element] (F-2-3) at (1, -3) {O};
            \node[element] (F-3-3) at (2, -3) {P};
            
            \pgfmathsetseed{41}
            \def\pointCount{5}
            
           	\foreach \x in {1,...,\pointCount} {
           		\foreach \y in {1,...,\pointCount} {
           			\pgfmathsetmacro{\xCoord}{\x+0.4*(rnd-0.5)}
           			\pgfmathsetmacro{\yCoord}{\y+0.4*(rnd-0.5)}
           			\coordinate (\x-\y) at (\xCoord-2.5, -\yCoord+1.5);
           		}
           	}
           	
           	\foreach \x in {1,...,\pointCount} {
           		\foreach \y in {1,...,\pointCount} {
           			\ifnum\x<\pointCount
           			\draw (\x-\y) -- (\the\numexpr\x+1\relax-\y);
           			\fi
           			\ifnum\y<\pointCount
           			\draw (\x-\y) -- (\x-\the\numexpr\y+1\relax);
           			\fi
           		}
           	}
        \end{scope}

        \begin{scope}[xshift=0cm]
        	 \def\textOffset{0.1}
            \node[anchor=center] at (0,1) {\large \textbf{BSP Schedule}};
            \node[bspElement] (G-0-0) at (0, -\textOffset) {A};
            \node[bspElement] (G-1-0) at (\scaleT, -1-\textOffset) {B};
            \node[bspElement] (G-2-0) at (2*\scaleT, -2-\textOffset) {C};
            \node[bspElement] (G-3-0) at (3*\scaleT, -3-\textOffset) {D};
            \node[bspElement] (G-0-1) at (-\scaleT, -1-\textOffset) {E};
            \node[bspElement] (G-1-1) at (0, -2-\textOffset) {F};
            \node[bspElement] (G-2-1) at (\scaleT, -3-\textOffset) {G};
            \node[bspElement] (G-3-1) at (2*\scaleT, -4-\textOffset) {H};
            \node[bspElement] (G-0-2) at (-2*\scaleT, -2-\textOffset) {I};
            \node[bspElement] (G-1-2) at (-\scaleT, -3-\textOffset) {J};
            \node[bspElement] (G-2-2) at (0, -4-\textOffset) {K};
            \node[bspElement] (G-3-2) at (\scaleT, -5-\textOffset) {L};
            \node[bspElement] (G-0-3) at (-3*\scaleT, -3-\textOffset) {M};
            \node[bspElement] (G-1-3) at (-2*\scaleT, -4-\textOffset) {N};
            \node[bspElement] (G-2-3) at (-\scaleT, -5-\textOffset) {O};
            \node[bspElement] (G-3-3) at (0, -6-\textOffset) {P};

            \foreach \i in {0, ..., 6} {
            	\node[anchor=north west] at (-3.8*\scaleT,0.05+\scaleBSP-\i*\scale) {\tiny Parallel For Loop - \textit{Bucket}};
                \draw[dashed] (-3.7*\scaleT,\scaleBSP-\i*\scale) -- (3.7*\scaleT,\scaleBSP-\i*\scale) -- (3.7*\scaleT,-\scaleBSP-\i*\scale) -- (-3.7*\scaleT,-\scaleBSP-\i*\scale) -- cycle;
            }
            \foreach \i in {0, ..., 5} {
                \draw[->, line width=0.4mm] (0, -\scaleBSP-\i*\scale) -- (0, \scaleBSP-\i*\scale-\scale);
            }
        \end{scope}

        \begin{scope}[xshift=5cm]
            \node[anchor=center] at (0, \scale) {\large \textbf{AMT Schedule}};
            \node[amtElement] (G-0-0) at (0, 0) {A};
            \node[amtElement] (G-1-0) at (\scaleT, -\scale) {B};
            \node[amtElement] (G-2-0) at (2*\scaleT, -2*\scale) {C};
            \node[amtElement] (G-3-0) at (3*\scaleT, -3*\scale) {D};
            \node[amtElement] (G-0-1) at (-\scaleT, -\scale) {E};
            \node[amtElement] (G-1-1) at (0, -2*\scale) {F};
            \node[amtElement] (G-2-1) at (\scaleT, -3*\scale) {G};
            \node[amtElement] (G-3-1) at (2*\scaleT, -4*\scale) {H};
            \node[amtElement] (G-0-2) at (-2*\scaleT, -2*\scale) {I};
            \node[amtElement] (G-1-2) at (-\scaleT, -3*\scale) {J};
            \node[amtElement] (G-2-2) at (0, -4*\scale) {K};
            \node[amtElement] (G-3-2) at (\scaleT, -5*\scale) {L};
            \node[amtElement] (G-0-3) at (-3*\scaleT, -3*\scale) {M};
            \node[amtElement] (G-1-3) at (-2*\scaleT, -4*\scale) {N};
            \node[amtElement] (G-2-3) at (-\scaleT, -5*\scale) {O};
            \node[amtElement] (G-3-3) at (0, -6*\scale) {P};

            \foreach \x in {0, 1, 2, 3} {
                \foreach \y in {0, 1, 2, 3} {
                    \ifnum\x<3
                        \draw[->, line width=0.4mm] (G-\x-\y) -- (G-\the\numexpr\x+1\relax-\y);
                    \fi
                    \ifnum\y<3
                        \draw[->, line width=0.4mm] (G-\x-\y) -- (G-\x-\the\numexpr\y+1\relax);
                    \fi
                }
            }
        \end{scope}
        
    \end{tikzpicture}
    \caption{Diagram showing how UnSNAP parallelizes the spatial domain for an angular sweep on an example mesh (left) with Bulk Synchronous Parallelism (centre) and Asynchronous Many-Task (right).}
    \label{fig:schemes}
    \end{center}
\end{figure}

Asynchronous Many-Task (AMT) is an alternative parallelism paradigm to BSP where tasks are defined as ``a sequence of instructions within a program that can be processed concurrently with other tasks in the same program'' \citep{Thoman2018}. These tasks are generated by application code and are scheduled onto threads by the programming model's runtime. The key advantage is that each task can specifically depend on other tasks, allowing data dependencies to be defined with much finer granularity, allowing for asynchronous execution. The differences between BSP and AMT are illustrated in figure \ref{fig:schemes}.

\subsection{Motivation}
\label{sec:motivation}

As described in section \ref{sec:par-strat}, UnSNAP is currently parallelized using BSP by creating a \textit{bucket} containing all elements within a t-level, bucket elements and energy groups are then computed simultaneously. One issue with this is that it creates a synchronization barrier at the end of every single t-level. This means that unless all cores within the processor complete their work at the same time, cores will sit idle while waiting at the barrier. Primarily, this could happen if the number of cores is not divisible by the number of elements multiplied by the number of energy groups. In modern processors it is common to have more cores than the number of energy groups and the number of elements within a bucket varies based on mesh data and sweep progression. As such, it is common for the work to not evenly divide by the number of cores, leaving a loop remainder.

\begin{figure}[htbp]
	\centering
	\includegraphics[width=0.95\linewidth]{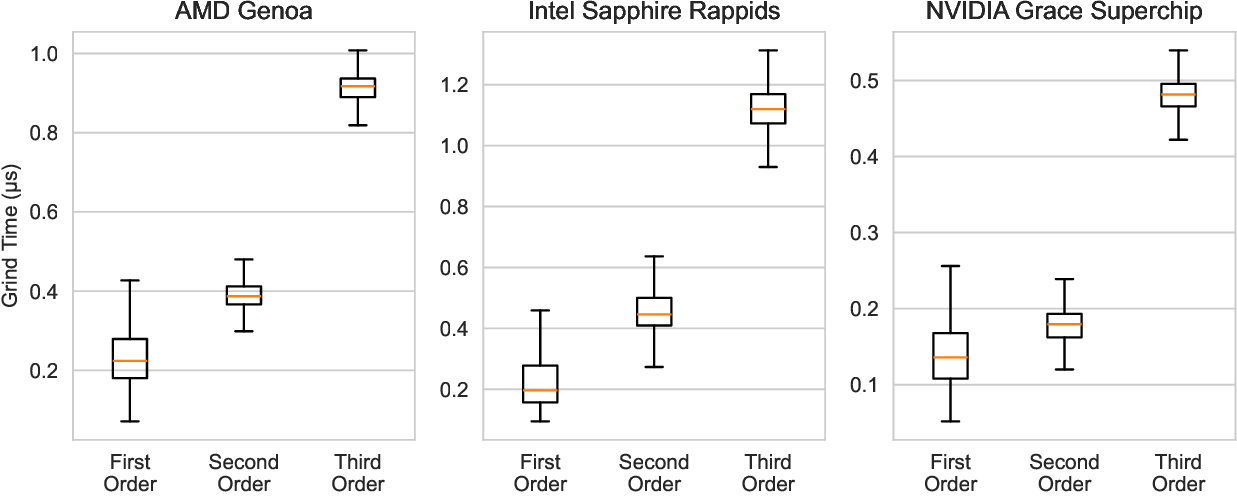}
	\caption{Distributions of compute time for a single unknown in the angular flux array (grind time). Metrics taken from a $16\times 16 \times 16$ grid with one angle per octant, 16 energy groups, one inner iteration, and one outer iteration. This totals 524288 grind times in each case.} 
	\label{fig:grind-time}
\end{figure}

Moreover, although the number of instructions within each FEM solve is the same, the number of cycles it takes to execute them can vary depending on cache utilization. The grind time is the time taken to update a single value of angular flux (i.e., all spatial degrees of freedom in a single spatial finite element). In figure \ref{fig:grind-time}, we show the distribution of grind times for first, second, and third orders on processors from three  hardware vendors. We see that there is a significant deviation of the distribution of grind times.  This is particularly clear for first order problems and becomes less prevalent for higher orders. This further leads to load imbalance, causing threads to wait longer at synchronization barriers before progressing. 

Other effects that could cause load imbalance and lead to a greater time spent at synchronization barriers are context switches to unrelated tasks and kernel operations. During profiling we see a small number of cycles are attributable to these operations. Theoretically, node instability and core defects could also cause load imbalances and therefore greater time spent at synchronization barriers.

We apply AMT to UnSNAP to remove synchronization barriers between each t-level and to leverage load balancing mechanisms, theoretically yielding greater processor utilization. However, in order for AMT to achieve higher performance than BSP, the methods used to generate tasks and schedule them onto processors must be as efficient as possible - in particular have less overhead and yield increased processor utilization than the BSP approach. AMT as a parallelism paradigm has been thoroughly studied and many programming frameworks and runtimes exist to support it \citep{Thoman2018}. Specifically, we choose the OpenMP programming model which has supported AMT since 2008 through the \texttt{task} clause with further features continually added in subsequent specifications \citep{OpenMP2008}.

One important semantic point is that we must be clear on the difference between task generation and scheduling. Generation\footnote{Also known in some literature as task \textit{spawning}.} concerns how the application creates new tasks and the structure of their dependencies to inform the runtime of the programming model. Scheduling concerns how the runtime of the programming model represents tasks and decides which thread will execute them and when. These topics are discussed in the following two sections.

\subsection{Generation}

There are two predominant methods of generating tasks: breadth-first generation or work-first generation. Breadth-first generation involves a single thread traversing the space of work and generating tasks for each work item without completing the work. Other threads can immediately begin executing tasks as they are generated. Work-first generation typically takes a recursive approach to creating tasks where a thread first executes the work of the task then recursively generates the next tasks. Within $S_N$ sweeps, the latter reduces the number of tasks simultaneously existing as they only exist at the wavefront. However, it increases the complexity of task generation as multiple threads can generate tasks at once needing careful control not to erroneously generate the same task multiple times or allow out of order execution.

We use a recursive work-first task generation approach. A task is first started for the first element in the sweep. This task then executes the work to assemble and solve the matrix system for the element to obtain an angular flux for each element node. This task then generates tasks for the downwind elements. We found that a work first approach significantly decreases the overheads of generating and scheduling tasks as there are fewer total tasks in existence at any given time and there is no need to use the task dependency system of the programming model's runtime as a custom approach can be implemented. 

\subsection{Scheduling}

Once tasks have been generated by the application, it is up to the runtime of the programming model to schedule these tasks onto processor threads. Runtimes typically implement this through one of two types of task queues, a single centralized queue or distributed queues where each processor core has a queue. We find that a centralized queue is not effective as it creates a single point of resource contention which all cores must attempt to access when dequeueing a task. With a greater core count, the chance for contention increases. As modern hardware trends favor increasing core count, this problem only becomes magnified.

Using distributed task queues not only decreases resource contention but allows improved cache locality through choice of task placement. Default behavior is to have the new task placed on the queue of the generating thread, though runtimes typically have mechanisms to allow the generating thread to choose which task queue to place the new task on. This allows the task to be placed on a queue of the thread which last accessed a specific piece of data which increases the chance for the data to still exist in a lower level cache. In particular, OpenMP has an \texttt{affinity} clause which we make use of in our implementation to place the task on the queue of the thread which last accessed a particular item within the angular flux array.

An additional mechanism of distributed queues is that they allow cache aware load balancing through work stealing. If a thread finishes executing a task and finds no further tasks within its own task queue, then it will take a task from another thread's task queue. The order in which it looks through other task queues is determined by the cache hierarchy. It will look through all task queues starting from the thread which has a common shared cache, starting at the lowest level and working up the hierarchy then looking first at local memory, then remote memory \citep{Klemm2021}. This allows for load balancing while minimizing slow fetches from higher latency cache and memory locations.

It should be noted that scheduling mechanisms are implementation specific so behavior will vary based on the OpenMP runtime used. The GNU OpenMP runtime uses a centralized queue, whereas the LLVM OpenMP runtime and its derivatives use distributed task queues. Therefore, to leverage the benefits of distributed queues, UnSNAP will hereinafter be compiled using the Cray Compiler Environment 17.0.1.

\subsection{Granularity}

Task granularity concerns choosing the size of tasks which execute the work of an entire problem. Larger tasks mean that there will be fewer tasks and vice versa. If the task size is too large, there could not be enough tasks to keep all cores busy leading to starvation. This complication has only become more prominent with the trend of increased core count in newer processors. Conversely, if the task size is too small then there will be a large number of tasks exceeding the parallelism of the processor and only incurring further scheduling overheads, such as resource contention, scheduling complexity, and increased memory requirements. This parameter is extremely difficult for AMT applications to fine-tune as it is problem and hardware-specific. Effectively, we want to ensure that there is always at least one task per thread while not introducing additional overheads.

\begin{figure}
	\centering
	\includegraphics[scale=0.8]{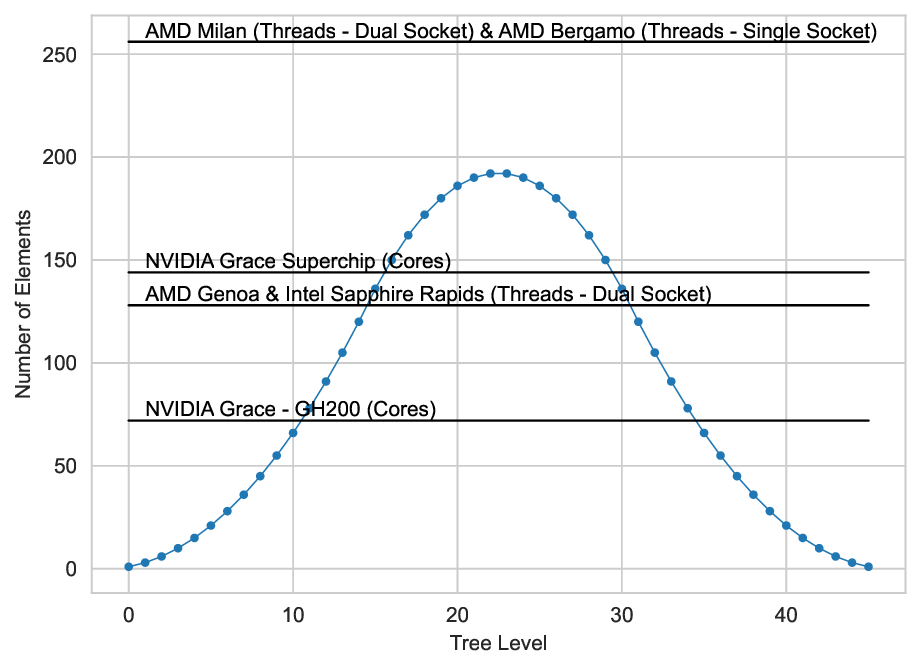}
	\caption{A comparison between the number of elements within each tree level and the number of processor threads/cores.} 
	\label{fig:parallelism}
\end{figure}

For $S_N$ sweeps we must expose enough parallelism while not creating more tasks than necessary. As all elements within a t-level do not depend on each other, they can be computed simultaneously. However, if each process only parallelizes over the element dimension of angular flux then not enough parallelism will be exposed for modern processors with high core counts. There are typically fewer elements at a sweep wavefront at any given time than the number of threads within modern processors, this is depicted in figure \ref{fig:parallelism}, where the number of elements within each t-level is compared to the number of cores within a variety of current CPUs. This comparison assumes the sweep is being performed on a $16\times16\times16$ grid of elements, in reality, the number of elements within each t-level will be data dependent on the unstructured grid layout. It will also depend on the chosen spatial decomposition and angular direction. Note too that the standard KBA spatial decomposition only increases one spatial dimension, and so the conclusions drawn from Figure 4 are applicable to any mesh 16x16xN under that regime. At the beginning and end of the sweep this issue is particularly prominent. Therefore, another dimension of the angular flux domain must be parallelized across. The BSP implementation of UnSNAP solves this problem by also parallelizing over energy groups and if the simultaneous angle scheme option is selected then also across angle. As long as there are enough energy groups and angles, then enough parallelism can be exposed for the vast majority of the grid, except the very beginning and end.

For an AMT algorithm to avoid starvation on modern high core count processors, it must likewise simultaneously compute along another dimension of the problem domain. In this context, the possible dimensions could be energy groups, angles, or both. If we chose to parallelize across just energy groups then there are two approaches: recursively sweep through the mesh for every energy group; or perform a single recursive sweep through the mesh but at each element generate a worker task for each energy group. The former approach is inefficient as each recursive sweep incurs a computational overhead while generating tasks. The latter is more efficient but still incurs additional overhead due to generating multiple tasks for each element. Alternatively, if we choose to additionally parallelize over angles, we can begin a recursive sweep for each angle simultaneously. This ensures the minimum number of tasks are generated while respecting sweep dependencies, minimizing the generation and scheduling overhead while exposing sufficient parallelism. This is the method we chose as it exposes enough parallelism while keeping the overheads to a minimum. Otherwise, we could additionally parallelize over both dimensions, simultaneously sweeping over all angles while creating a task for each energy group. However, we found that this method was slower due to increased memory resources and scheduling overhead required.

Another middle ground approach to exposing more parallelism is to arrange energy groups and angles into sets. Each set could be computed simultaneously and independently from each other to partially increase parallelism across these dimensions. This approach was introduced by \citet{Kunen2015} in the KRIPKE mini-app where these sets are known as GroupSets and DirectionSets. However, again, this approach was found to be slower due to additional sweeps required and increased overhead on the tasking runtime.

\subsection{Implementation Summary}

Taking the outcomes of understanding task generation, scheduling and granularity, our new algorithm works as follows. We begin by simultaneously generating a task for the first element in the sweep for every angle within an octant. Each task performs the assemble solve for all energy groups of that element before recursively generating tasks for any downwind elements where all other dependencies are satisfied. To do this, each element has a counter which represents the number of times an upwind element has been executed, this is incremented atomically when the upwind element's task executes. The counter allows each upwind neighbor to check if it is the last dependency to execute for a given element within a sweep to ensure that only one task schedules the downwind task. 

This bespoke system of atomic counters is used instead of the OpenMP \texttt{depend} clause as it is incapable of handling registering dependencies on tasks which do not yet exist. If an upwind task tries to schedule a downwind task and register a dependence on a different task which has not yet been generated then that downwind task will assume the dependence to be trivially satisfied. As tasks execute in an asynchronous non-deterministic order, this scenario can occur and lead to out of order execution of tasks and incorrect results. 

This process is repeated for all octants, inner, and outer iterations, respecting the mini-app's design constraint to not parallelize over octants. Vectorization remains the same as the previous implementation, with loops within the FEM solve being vectorized. Additionally, we ensure that a programming model with a runtime that implements a distributed task queue is used. This implementation is summarized in pseudo-code in algorithm \ref{alg:recursive}.

\begin{subequations}  \label{eq:pop}
	\begin{align}
		integrated\_flux(E) = \sum_n \phi(\overset\rightarrow r_n,E) \label{eq:pop_e} \\
		total\_integrated\_flux = \sum_n integrated\_flux(E_n) \label{eq:pop_tot}
	\end{align}
\end{subequations}

In equation \ref{eq:pop_e} we define the \textit{integrated flux} as the summation of scalar flux values over the spatial dimension which produces a value for each energy group. Furthermore, we define the \textit{total integrated flux} in equation \ref{eq:pop_tot} as the summation of the integrated flux values over all energy groups to produce a single value. We compare the integrated flux values at the end of each outer iteration and the final total integrated flux value from both the new AMT and existing BSP implementations. If the values have not deviated greater than would be expected from changes to the floating-point arithmetic order of operation then we know that the new implementation has not altered the result of the simulation.

\begin{algorithm}[htbp]
	\caption{Recursive work-first AMT method for unstructured grid sweeps}\label{alg:recursive}
	\begin{center}
		\begin{algorithmic}
			\Require A set of energy groups $E$ and a set of angles $\hat\Omega$.
			\Procedure{RecursiveSweep}{$elem, a, E$}
			\ForAll{$e \in E$} \Comment{Iterate energy groups sequentially}
			\State $\psi(elem,a,e) \gets \textproc{Solve}({elem,a,e})$
			\EndFor
			\State $N \gets \textit{Downwind neighbours of } elem$
			\ForAll{$n \in N$}  \Comment{Iterate neighbors sequentially}
			\If{$elem$ is the final upwind dependency of $n$ }
			\State Generate \textproc{RecursiveSweep}($n, a, E$) as a Task
			\EndIf
			\EndFor
			\EndProcedure
			\\
			\ForAll{$a \in \hat\Omega$} \Comment{Iterate angles simultaniously}
			\State $elem_0 \gets \textit{The first element in the sweep}$
			\State Generate \textproc{RecursiveSweep}($elem_0, a, E$) as a Task
			\EndFor
		\end{algorithmic}
	\end{center}
\end{algorithm}

\subsection{Numerical Results}

\begin{figure}[htbp]
\centering
\includegraphics[width=0.95\linewidth]{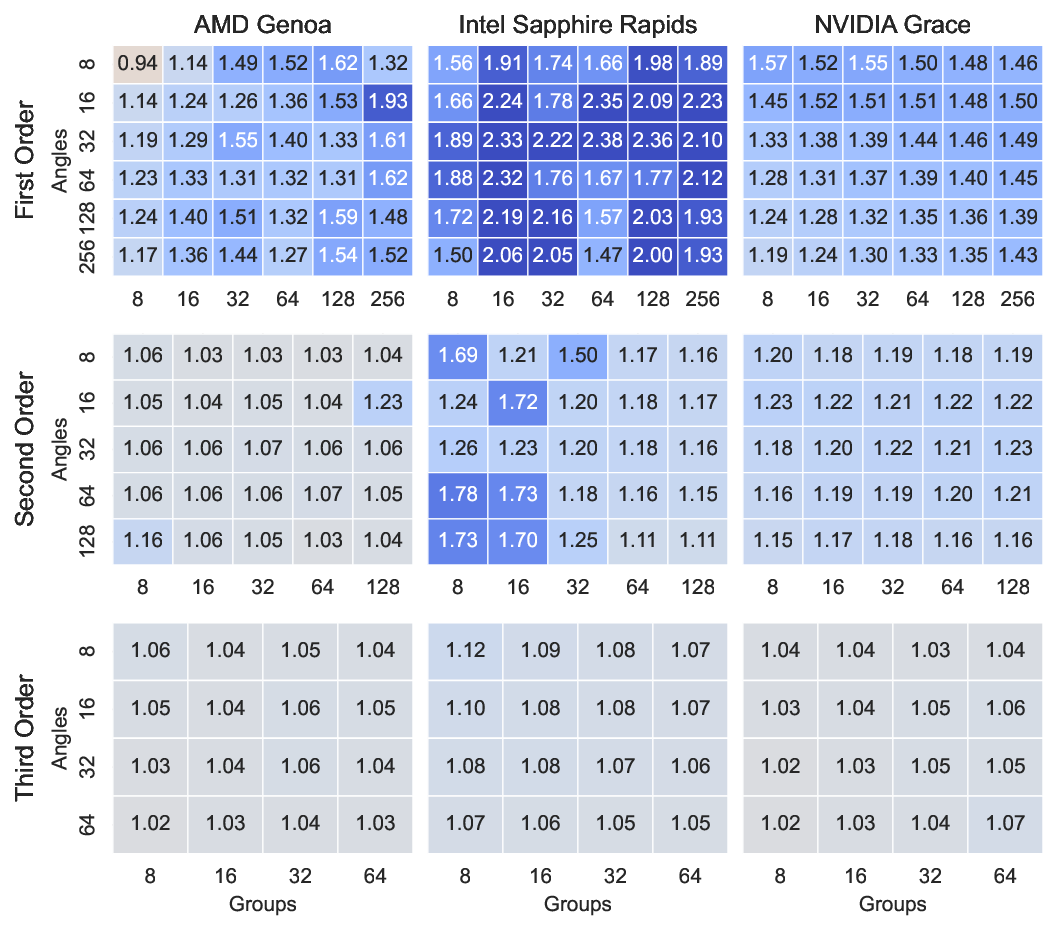}
\caption{Speedups between the previous BSP method and new AMT method for UnSNAP across compute nodes from three different 
 hardware vendors.} 
 \label{fig:heatmap}
\end{figure}

To test the new implementation, we ran benchmarks for both the BSP and AMT versions while using a $16\times 16\times 16$ grid of elements, with five outer and five inner iterations while enabling parallelization across angle. As the size of the input problem impacts performance, we run it across a variety of angle and group sizes and for first, second and third order finite element solves. In each case, the largest input size for each order was determined by the maximum amount of memory available on the systems. As with the BSP benchmarks, where SMT is used, we bind threads to physical threads and otherwise we bind them to cores. We run these results using an AMD Genoa, Intel Sapphire Rapids and NVIDIA Grace Superchip CPUs, as summarized in table \ref{tab:compute-nodes}. These results are summarized in figure \ref{fig:heatmap} which shows the speedup of the new AMT method over the previous BSP approach. Corresponding runtimes for 16 energy groups and 16 angles are also shown in table \ref{tab:times}. 

\begin{table}[htbp]
	\tbl{Runtimes of BSP and AMT UnSNAP implementations. Metrics taken from a $16\times 16 \times 16$ grid with 16 angles per octant, 16 energy groups, five inner iterations, and five outer iterations.}{
		\begin{tabular}{lcccccc} \toprule
			\multirow{3}{*}{\textbf{Architecture}} & \multicolumn{6}{c}{\textbf{Runtime (s)}} \\
			& \multicolumn{2}{c}{First Order} & \multicolumn{2}{c}{Second Order} & \multicolumn{2}{c}{Third Order} \\
			& BSP & AMT & BSP & AMT & BSP & AMT \\ \midrule
			AMD Genoa & 2.42 & 1.85  & 21.47 & 20.45 & 83.23 & 80.15 \\ \midrule
			Intel Sapphire Rapids & 4.83 & 1.99 & 26.52 & 15.15 & 98.02 & 90.30  \\ \midrule
			NVIDIA Grace Superchip & 1.75 & 1.09 & 7.22 & 5.87 & 41.10 & 39.48 \\ \bottomrule
	\end{tabular}}
	\label{tab:times}
\end{table}

\begin{figure}
	\centering
	\includegraphics[scale=0.7]{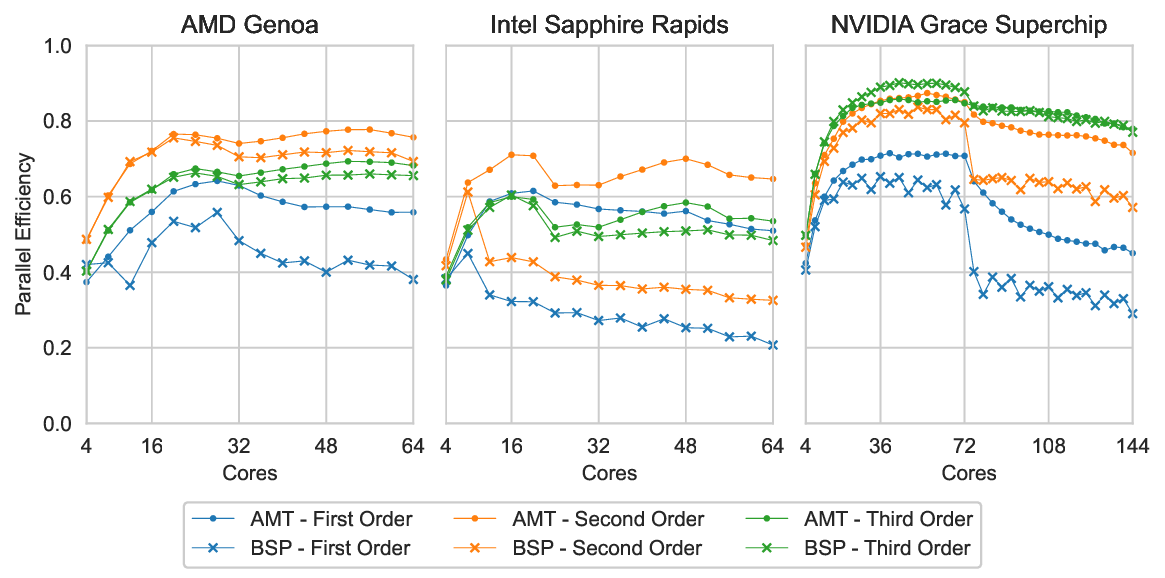}
	\caption{Comparison of the parallel efficiency of UnSNAP between the AMT and BSP implementations.} 
	\label{fig:scaling}
\end{figure}

In almost all cases, the new method matches or outperforms the previous. AMT particularly shows performance improvements in first order FEM solves, and shows diminishing improvements with increasingly higher FEM orders. Additionally, we observe that on the NVIDIA Grace Superchip, smaller problems afford a more significant speedup. Figure \ref{fig:scaling} shows the parallel efficiency of both methods across first, second and third FEM orders. Here we see that the new method generally achieves a greater parallel efficiency as the problem scales compared to the previous method. Additionally, we can see that as Grace scales to use the first socket in the chip, it achieves nearly 90\% parallel efficiency for third order problems before seeing a decrease as the program begins to utilize the second socket.

\subsubsection{Synchronisation}

As discussed in section \ref{sec:motivation}, BSP approaches typically work well when there is a uniform number of work items assigned to each processor core and each work item is uniform. If this is the case, processor cores will complete their assigned work in roughly the same amount of time, minimizing the time that cores sit idle at the barrier waiting to synchronize. In the case of UnSNAP, as there will be a different number of items within each bucket depending on the input mesh, it cannot be guaranteed that there will be an even distribution of work divided across all cores. Additionally, although each assemble solve compiles to the same number of instructions, it is not guaranteed that they will be executed in the same number of processor cycles due to level of cache or memory that the core must fetch the required data from not always being the same, this is depicted in figure \ref{fig:grind-time}. These two factors lead to an imbalanced workload within parallel for code blocks, causing processors to sit idle at synchronization barriers. For the BSP implementation, we measure that for first, second and third order problem sizes the number of cycles where cores are waiting at synchronization locks is 33.58\%, 31.83\%, and 8.36\% respectively\footnote{Metrics taken from a $16\times 16 \times 16$ grid with 32 angles, 32 energy groups, 5 inner iterations, 5 outer iterations, and a parallel angle schedule on Intel Sapphire Rapids.}; a significant portion of the program runtime. The load balancing from work stealing within an AMT approach means that there are no global synchronization barriers (which occurred at the end of each t-level/wavefront in the BSP approach). There are however some synchronization locks required for handling of shared resources related to task scheduling, however, processor cycles spent waiting at locks still reduces to 9.45\%, 3.77\%, and 2.11\% for first, second and third order problems respectively of the total AMT runtime. However, there are other overheads introduced in the AMT implementation attributable to task orchestration, of processor cycles, these make up 12.06\%, 3.06\%, and 2.19\% for first, second and third order problems respectively.\footnote{For the BSP implementation the approximate total cycles which occurred during this benchmark were: $2.23\times 10^{12}$, $1.06\times 10^{13}$, and $3.90\times 10^{13}$ for each order respectively. For the AMT the approximate total cycles which occurred were: $9.19\times 10^{11}$, $6.00\times 10^{12}$, and $3.51\times 10^{13}$ for each order respectively.} Overall, the new AMT implementation shows a significant decrease in synchronization time despite the additional overheads introduced. This allows for a greater portion of processor time to be spent computing the assemble solve of the small matrix within the FEM solve.

\subsubsection{Cache Utilisation}

Table \ref{tab:cache} shows the cache hit rate for both AMT and BSP implementations across different hardware vendors. A high cache rate indicates that there is a high level of data reuse in the cache, and is something that should indicate high performance for a code with high levels of access to small data structures as is the case for FEM codes. Due to the size of the angular flux array, if it is accessed with low temporality, this causes challenges to the cache behavior \citep{Deakin2017}. In the first order case, we see that for both implementations the hit rate of L1 and L2 caches is still high with a decrease in hit rate for last level cache. The new AMT implementation has a similar hit rate for L1 cache, improves upon cache hit rate for L2 cache particularly on Sapphire Rapids, but can have decreased performance on L3 caches. On second and third order problems, we see that cache utilization of L1 and L2 cache very similar between implementations, however, there is a decrease in the hit rate for last level cache. This decrease can be particularly significant for Sapphire Rapids at almost 50\%, however, this does not appear to outweigh the benefits of AMT as there is no decrease in performance for third order problems. 

\setcounter{table}{1}
\begin{sidewaystable}\tbl{Comparison between BSP and AMT cache utilisation for UnSNAP. Metrics taken from a $16\times 16 \times 16$ grid with 16 angles per octant, 16 energy groups, five inner iterations, and five outer iterations.}{
	\begin{tabular}{lcccccccccc} \toprule
		\multirow{3}{*}{\textbf{Architecture}} & \multirow{3}{*}{\textbf{Cache Level}} & \multicolumn{9}{c}{\textbf{Cache Hit Rate (\%)}} \\
		& & \multicolumn{3}{c}{First Order} & \multicolumn{3}{c}{Second Order} & \multicolumn{3}{c}{Third Order} \\
		& & BSP & AMT & Difference & BSP & AMT & Difference & BSP & AMT & Difference \\ \midrule
		\multirow{4}{*}{AMD Genoa} & L1 & 98.79 & 98.45 & -0.34 & 97.76 & 97.70 & -0.07 & 90.31 & 90.03 & 	-0.28 \\
		& L2 & 85.15 & 92.03 & 6.88 & 96.93 & 97.54 & 0.61 & 99.64 & 99.70 & 0.06  \\
		& Local L3\textsuperscript{a} & 93.80 & 96.06 & 2.26 & 97.85 & 98.32 & 0.47 & 99.68 & 99.76 & 0.08 \\
		& Remote L3 & 55.61 & 50.46 & -5.14 & 22.98 & 17.16 & -5.82 & 22.01 & 15.59 & -6.42 \\ \midrule
		\multirow{3}{*}{Intel Sapphire Rapids} & L1 & 99.84 & 99.78 & -0.06 & 99.10 & 98.87 & -0.23 & 95.82 & 96.03 & 0.21 \\
		& L2 & 70.43 & 87.10 & 16.67 & 97.11 & 97.65 & 0.54 & 99.50 & 99.19 & -0.32  \\
		& L3 & 62.28 & 77.69 & 15.41 & 58.75 & 54.24 & -4.51 & 64.74 & 15.37 & -49.36 \\ \midrule
		\multirow{3}{*}{NVIDIA Grace Superchip} & L1 & 99.43 & 99.15 & -0.27 & 99.28 & 99.18 & -0.10 &	99.13 &	99.10 & -0.03 \\
		& L2 & 92.13 & 93.27 & 1.14 & 98.44 & 98.74 & 0.30 & 99.70 & 99.71 & 0.01 \\
		& L3 & 70.75 &	65.25 &	-5.51 &	70.75 &	61.64 &	-9.10 &	70.55 &	58.55 &	-12.00 \\ \bottomrule
	\end{tabular}}
	\tabnote{\textsuperscript{a}Unlike Sapphire Rapids and Grace, Genoa has a distributed L3 cache where each Core CompleX (CCX) has an L3 cache shared between four cores.}
	\label{tab:cache}
\end{sidewaystable}

As the size of lower level caches has increased, we see that their hit rate has increased in turn. It is now the case that the small matrix for each FEM solve can fit within a core's L1 cache even up to third order problems for the hardware used. This significantly, decreases the chance that it could be evicted from this cache. The per core cache capacity is highest for Grace where the small matrix for third order will only take up ~50\% of a core's L1 cache. This is reflected in the fact that we see little decrease in the cache hit rate from second to third order problems for both AMT and BSP implementations. These conclusions support the findings of previous work on cache utilization for FEM solves in sweeps where the application performance is either bound by the bandwidth of the cache which the small matrix fits within or main memory bandwidth from the data streamed while assembling the small matrix. This depends on the size of problem dimensions \citep{Deakin2020, Deakin2020MCHPC}.

\section{Conclusions}
\label{sec:conclusion}

We have created an updated review of the existing BSP algorithm for UnSNAP and presented its performance across a range or modern processors. Here we see that modern CPUs with an increased number of cores continue to improve the performance of the mini-app, particularly for third-order problems. Additionally, due to an optimized CUDA kernel for first-order problems, performance of the GPU code matches or outperforms CPU performance.

Secondly, we have described a new algorithm for parallelizing sweeps on unstructured grids which uses an AMT paradigm. In almost all cases this shows significant performance improvements over the previous method, particularly for lower orders. As the spatial finite element order increases, the significance of efficient generation and scheduling of tasks diminishes, however, this may show additional benefits in domain decompositions where wavefront parallelism alone is limited.

As results on the BSP version of UnSNAP show that GPUs can outperform CPUs for some input sizes, we aim to also make an implementation for the AMT algorithm which runs on GPUs. Specifically, we are working on writing an implementation using the CUDASTF library to target NVIDIA GPUs and to evaluate its performance against our AMT CPU implementation \citep{Augonnet2024}. 

Additionally, future work can be undertaken to widen the parallelization and communication schemes that the AMT algorithm implements. Section \ref{sec:par-strat} discussed the advantages and disadvantages of single and parallel angle parallelization schemes with further sections discussing why a parallel angle scheme is the only performant method of structuring an AMT algorithm. We aim to revisit a single angle parallelization scheme for AMT to compare approaches. Section \ref{sec:distributed} discussed other distributed memory communication patterns. Currently the AMT algorithm is only implemented for PBJ communication, however, we aim to implement a version for FPS communication. The increased granularity of work and load balancing within AMT algorithms should allow granular FPS communication allowing dependencies to be communicated sooner and more efficiently, leading to increased node utilization and a reduction in the start up time of the sweep.

\section{Funding}

The authors acknowledge the use of resources provided by the Isambard 3 Tier-2 HPC Facility. Isambard 3 is hosted by the University of Bristol and operated by the GW4 Alliance (https://gw4.ac.uk) and is funded by UK Research and Innovation; and the Engineering and Physical Sciences Research Council [EP/X039137/1]. The authors acknowledge the use of Blue Crystal Phase 4 HPC resource operated by the Advanced Computing Research Centre (ACRC) at the University of Bristol. This work makes use of the HPC Zoo compute cluster operated by the High Performance Computing Research Group at the University of Bristol. This work was funded by AWE PLC’s Future Technologies Programme.

\section{Disclosure statement}

No potential conflict of interest was reported by the authors.

\bibliographystyle{tfcad}
\bibliography{sources}

\end{document}